\documentclass[
reprint,amsmath,amssymb,aps,pra,floatfix,]{revtex4-1}
\usepackage{graphicx}	
\usepackage{dcolumn}	
\usepackage{bm}		
\usepackage{color}

\begin{document}

\title{A tabulation of the bound-state energies of atomic hydrogen}
\author{M. Horbatsch and E.A. Hessels}
\email{hessels@yorku.ca}

\affiliation{%
 Department of Physics and Astronomy, York University, Toronto, Ontario M3J 1P3, Canada
}%

\date{\today}

\begin{abstract}
We present tables for the bound-state energies for atomic hydrogen. 
The tabulated energies include the hyperfine structure, 
and thus this work extends the 
work of Rev. Mod. Phys. {\bf 84}, 1527 (2012), 
which excludes hyperfine structure. 
The tabulation includes corrections of the hyperfine structure due to the 
anomalous moment of the electron, due to the finite
mass of the proton, and due to off-diagonal matrix elements of 
the hyperfine Hamiltonian. These corrections are 
treated incorrectly in most other works. 
Simple formulas valid for all quantum numbers are 
presented for the hyperfine corrections. 
The tabulated energies have uncertainties 
of less than 1 kHz for all states.  
This accuracy is possible because of the 
recent precision measurement 
[Nature, {\bf 466}, 213 (2010); 
Science, {\bf 339}, 417] 
of the proton radius. 
The effect of this new radius on the energy levels is also tabulated,
and the energies are compared to precision measurements 
of atomic hydrogen energy intervals.
\begin{description}
\item[PACS numbers]
\verb+\pacs{31.15.ac,31.15.1j,06.20.Jr,12.20.-m,14.20.Dh}+
\end{description}
\end{abstract}

\maketitle

\section{introduction}
\label{sec:introduction}

The current state of the theoretical knowledge of the 
bound-state energy levels of atomic hydrogen 
has been presented in the CODATA analysis of 
Ref.~\cite{mohr2012codata}.
Here we expand on that work, by also including the 
hyperfine structure. Because of a recent precise determination 
\cite{antognini2013proton, pohl2010size}
of the rms proton charge radius 
$R_{\rm p}$,
obtained from
measurements of the $n$=2 intervals in muonic hydrogen, 
all of the atomic
hydrogen energy levels can now be determined to an accuracy 
of better than 1 kHz. 
In this work, 
we present tables for the energies
$E_{n\ell jf}$ of $|nljfm_{\!f}\rangle$ bound states of atomic hydrogen.  
These tables also indicate how the energy of each 
state is affected by the new value of $R_{\rm p}$ and
by the new determination of the Rydberg constant that results from 
this new $R_{\rm p}$ value.

The present work was performed because of the need for precision 
energy levels for the analysis of an ongoing measurement of the 
hydrogen $n$=2 Lamb shift, for which the hyperfine structure and
fine structure must be carefully considered. 
The tabulation is mostly based on the theoretical
and experimental work of others, 
and is intended as a practical
resource. 
   
\section{Overview}
\label{sec:overview}

To date, 
only three intervals in atomic hydrogen have been measured 
to 
an accuracy of better than 1 kHz:
\begin{eqnarray}
&&\nu(1{\rm S}_{1_{\!}/_{\!}2} f\!=\!1\!\to\!2{\rm S}_{1_{\!}/_{\!}2} f\!=\!1)\!:
2\,466\,061\,102\,474.806(10)\  {\rm kHz}, 
\nonumber \\
&&\nu(1{\rm S}_{1_{\!}/_{\!}2} f\!=\!0\!\to\!1{\rm S}_{1_{\!}/_{\!}2} f\!=\!1)\!:
1\,420\,405.751\,768(1)\  {\rm kHz}, 
\nonumber \\
&&\nu(2{\rm S}_{1_{\!}/_{\!}2} f\!=\!0\!\to\!2{\rm S}_{1_{\!}/_{\!}2} f\!=\!1)\!:
177\,556\,.8343(67)\  {\rm kHz},
\label{eq:Hexpt}
\end{eqnarray}
where the first and the last values were measured 
\cite{parthey2011improved,kolachevsky2009measurement}
by the 
H{\"a}nsch group, and the $1{\rm S}_{1_{\!}/_{\!}2}$ 
hyperfine interval is based on 
an analysis by Karshenboim
\cite{karshenboim2005precision} of 
all measurements of this interval.
Theoretical predictions are needed to determine all other intervals
to an accuracy of $<$1 kHz. 
In addition,  
precise values of fundamental constants are required.
The most important constant needed is the Rydberg constant 
($R_{\infty}$).
As will be discussed in Section~\ref{sec:gross},
this constant can be obtained from 
a combination of the theoretical predictions for the 
hydrogen atom
and the measurements of 
Eq.~(\ref{eq:Hexpt}). 

The second most important constant is the fine-structure constant ($\alpha$),
the value of which is known from a comparison between theory and 
measurements of the magnetic moment of the electron. For this work,
we use the CODATA 2014 value of 
\begin{equation}
\alpha=1/137.035\,999\,139(31),
\label{eq:alpha}
\end{equation}
which is derived almost entirely from the electron magnetic moment.
The proton's mass, magnetic moment and rms charge ratio are also needed.
The CODATA 2014 values for these quantities are
\begin{eqnarray}
m_{\rm p}/m_{\rm e} &=& 1836.152\,673\,89(17),
\label{eq:mpCODATA}
 \\
g_{\rm p}&=&5.585\,694\,702(17),
\label{eq:gpCODATA}
 \\
{\rm and}\  R_{\rm p}&=&0.8751(61)\  {\rm fm}.
\label{eq:RpCODATA}
\end{eqnarray}
A more precise determination of the rms charge radius has
been obtained using measurements 
\cite{pohl2010size,antognini2013proton} 
of the $n$=2 levels of muonic
hydrogen by the CREMA collaboration:
\begin{eqnarray}
R_{\rm p}&=&0.840\,87(39)\  {\rm fm}.
\label{eq:crema}
\end{eqnarray}
This value differs significantly from the CODATA 2014 determination.
We use this more precise determination of $R_{\rm p}$ 
in this work, but 
also tabulate the shifts in the bound-state hydrogen energies 
that would result if the CODATA 2014 value were used. 

The other two required constants are 
the Compton wavelength, $\lambda_C$, 
and the muon mass, $m_{\mu}$, which
have CODATA 2014 values of
\begin{eqnarray}
\mathchar'26\mkern-9mu\lambda_C=\frac{\lambda_C}{2\pi}
&=&386.159\,267\,64(18)\ {\rm fm},
\label{eq:lambdaCCODATA} \\
{\rm and}\ m_{\mu}/m_{\rm e} &=& 206.768\,2826(46).
\label{eq:mmuCODATA}
\end{eqnarray}

The constants of 
Eqs.~(\ref{eq:alpha}),~(\ref{eq:mpCODATA}),~(\ref{eq:gpCODATA}),~(\ref{eq:crema}),~(\ref{eq:lambdaCCODATA}),~and~(\ref{eq:mmuCODATA})
are all determined
using physical systems other than atomic hydrogen, and therefore
should be considered to be external inputs to the theory.
Eqs.~(\ref{eq:alpha}),~(\ref{eq:mpCODATA}),~(\ref{eq:gpCODATA})~(\ref{eq:lambdaCCODATA}),~and~(\ref{eq:mmuCODATA}) are known with sufficient precision
as to lead to uncertainties in the hydrogen energies of 
much less than 0.1~kHz. The proton charge radius $R_{\rm p}$
of Eq.~(\ref{eq:crema}), however, despite its more precise 
determination, is still one of the leading causes for uncertainty
in the hydrogen energies.  

The binding energy of an $|n\ell jfm_{\!f}\rangle$ state of hydrogen can be 
written as 
\begin{equation}
E_{n\ell jf}=
E_n^{\rm (g)}
+E^{\rm (fs)}_{n\ell j}
+E^{\rm (hfs)}_{n\ell jf},
\label{eq:Henergy}
\end{equation}
where 
\begin{equation}
E_n^{\rm (g)}=-\frac{\mathcal{R}}{n^2}\frac{m_{\rm r}}{m_{\rm e}}
\label{eq:Eg}
\end{equation}
is the gross structure,
$E^{\rm (fs)}_{n\ell j}$ is the fine structure contribution, 
and $E^{\rm (hfs)}_{n\ell jf}$ is the hyperfine structure contribution.
The energies are, of course, independent of $m_{\!f}$ in the absence
of an external field.
Here, 
$\mathcal{R}$=$c R_{\infty}$,
$c$ is the speed of light,
$R_{\infty}$ is the Rydberg constant, 
$m_{\rm r}$=$m_{\rm e}$$m_{\rm p}$/$M$ is the reduced mass, 
$M$=$m_{\rm e}$+$m_{\rm p}$,
and
$m_{\rm e}$ and $m_{\rm p}$ are the electron and proton masses,
respectively, 
For this work, Planck's constant $h$ is suppressed, and
all energies are given in frequency units.

\section{fine-structure energy contributions}
\label{sec:fs}

The values of $E^{\rm (fs)}_{n\ell j}$ are discussed in detail in 
Ref.~\cite{mohr2012codata}:
\begin{eqnarray}
E^{\rm (fs)}_{n\ell j}&=&\!
\Delta E_{M}
\!+\!E_{\rm S}
\!+\!E_{\rm R}
\!+\!E_{\rm SE}^{(2)}
\!+\!E_{\rm VP}^{(2)}
\nonumber \\
\!&+&\!E^{(4)}
\!\!+\!E^{(6)}
\!+\!E_{\rm RR}
\!+\!E_{\rm SEN}
\!\!+\!E_{\rm NS},
\label{eq:Efs}
\end{eqnarray}
where $\Delta E_{M}$ gives the mass-corrected Dirac fine structure, 
$E_{\rm S}$ and $E_{\rm R}$ are relativistic recoil corrections, 
$E_{\rm SE}^{(2)}$ and $E_{\rm VP}^{(2)}$ are self-energy 
and vacuum polarization quantum-electrodynamics (QED)
corrections, $E^{(4)}$ and $E^{(6)}$ are higher-order QED
corrections, $E_{\rm RR}$ and $E_{\rm SEN}$ are small QED 
corrections due to the finite mass of the proton, 
and $E_{\rm NS}$ is the correction for the size 
(rms charge radius) of 
proton. For completeness, we include the formulas given in
Ref.~\cite{mohr2012codata}
that give contributions which are significant to this work,
leaving out energy terms and corrections that lead
to contributions of $<$0.025 kHz 
to our tabulated values. 
Because of the precisely-measured 1S-2S interval of
Eq.~(\ref{eq:Hexpt}), $E^{\rm (fs)}$ for the 1S state energy
needs to be determined less accurately, and is only necessary 
for the determination of the Rydberg constant, as will be 
discussed in Section~\ref{sec:gross}.     

From Ref.~\cite{mohr2012codata}, we obtain: 
\begin{eqnarray}
\Delta E_M\!&=&\!
E_M\!-\!Mc^2\!-\!E^{\rm (g)}_n\!=\!
\frac{
2[f_{n j}\!-\!1]
-[f_{n j}\!-\!1]^2\frac{m_{\rm r}}{M } 
}
{\alpha^2 } 
\frac{\mathcal{R}m_{\rm r}}{m_{\rm e}}
\nonumber \\
&&+\frac{\mathcal{R}}{n^2}\frac{m_{\rm r}}{m_{\rm e}}
+
\frac{1\!-\!\delta_{\ell 0}}{2 \ell\!+\!1}
\frac{\alpha^2 \mathcal{R} m_{\rm e}^2}{\kappa_{\ell j} n^3 m_{\rm p}^2},
\label{eq:EM}
\end{eqnarray}
\begin{eqnarray}
E_{\rm S}\!=\!
\frac{2m_{\rm r}^3\alpha^3 \mathcal{R}}{m_{\rm e}^2 m_{\rm p}\pi n^3}
\Big\{
\frac{\!\delta_{\ell 0}}{3}\lambda
\!-\!\frac{8}{3} \beta_{n \ell}
\!-\!\frac{\delta_{\ell 0}}{9} 
\!-\!\frac{7}{3} a_{n\ell}
\!-\! 2 \delta_{\ell 0} \frac{m_{\rm e}}{m_{\rm p}}
\Big\},\ \ \ \ \ \ \ 
\label{eq:ES}
\end{eqnarray}
\begin{eqnarray}
E_{\rm R}\!&=&\!
\frac{2
m_{\rm e}
\alpha^4 
\mathcal{R}}{
m_{\rm p}
\pi
 n^3}
\Big\{
D_{60}^{(n\ell)}
\!-\!\frac{11 \delta_{\ell 0}}
{60 }
\alpha
\lambda^2
\!+\!D_{71}^{(n \ell j)} \alpha
\lambda
\Big\},\ \ \ \ \ \ \ 
\label{eq:ER}
\end{eqnarray}
\begin{eqnarray}
E_{\rm SE}^{(2)}\!=\!
\frac{2m_{\rm r}^3\alpha^3 \mathcal{R}}{m_{\rm e}^3 \pi n^3}
\Big\{&&
\frac{4 \delta_{\ell 0}}{3}\lambda
\!-\!\frac{4\beta_{n \ell}}{3}
\!+\!\frac{10\delta_{\ell 0}}{9}
\!-\!\frac{(1\!-\!\delta_{\ell 0})}
{2\kappa_{\ell j}(2\ell\!+\!1)}\frac{m_{\rm e}}{m_{\rm r}}
\nonumber \\
&&
\!+\big(\frac{139}{32}\!-\!2 \ln 2\big)\pi\delta_{\ell 0}\alpha
\!-\!\delta_{\ell 0} \alpha^2 \lambda^2
\nonumber \\
&&
\!+A_{61}^{(n \ell j)} \alpha^2 \lambda
\!+\!G_{\rm SE}^{(n \ell j)} \alpha^2
\Big\},
\label{eq:E2SE}
\end{eqnarray}
\begin{eqnarray}
E_{\rm VP}^{(2)}\!=
\frac{2m_{\rm r}^3\alpha^3 \mathcal{R}}{m_{\rm e}^3 \pi n^3}
&&\Big\{
\Big[
\!-\!\frac{4}{15}\big(1\!+1.67\frac{m_{\rm e}^2}{m_{\mu}^2}\big)
\!+\!\frac{5\pi}{48}\alpha
\!-\!\frac{2}{15} \alpha^2 \lambda
\nonumber \\
&&
\!+\big(\frac{19}{45}\!-\!\frac{\pi^2}{27}\big)\alpha^2
\Big]
\delta_{\ell 0}
\!+\! G_{\rm VP}^{(1)(n \ell j) } \alpha^2
\Big\},\ \ \ \ 
\label{eq:E2VP}
\end{eqnarray}
\begin{eqnarray}
E^{(4)}\!=\!
\frac{2m_{\rm r}^3\alpha^4\mathcal{R}}{  m_{\rm e}^3 \pi^2 n^3}
&&\Big\{
0.53894 \delta_{\ell 0}\!+\!\frac{0.3285 (1\!-\!\delta_{\ell 0})}
{\kappa_{\ell j}(2\ell\!+\!1)}
\frac{m_{\rm e}}{m_{\rm r}}
\nonumber \\
&&-21.554  \delta_{\ell 0} \alpha
\!-\!\frac{8\delta_{\ell 0}}{27}\alpha^2 \lambda^3
\!+\! B_{62}^{(n\ell)} \alpha^2 \lambda^2
\nonumber \\
&&+B_{61}^{(n \ell j)}\alpha^2 \lambda
\!+\! B_{60}^{(n \ell j)} \alpha^2
\Big\},
\label{eq:E4}
\end{eqnarray}
\begin{eqnarray}
&&E^{(6)}\!\!=\!\!
\frac{2m_{\rm r}^3\alpha^5\mathcal{R}}{m_{\rm e}^3  \pi^3 n^3}
\Big\{
0.418 \delta_{\ell 0}
\!-\!\frac{1.2(1\!-\!\delta_{\ell 0})}
{\kappa_{\ell j}(2\ell\!+\!1)}\frac{m_{\rm e}}{m_{\rm r}}
\!+\!C_{50}\alpha
\Big\},\ \ \ \ \  
\label{eq:E6}
\end{eqnarray}
\begin{eqnarray}
&&E_{\rm NS}\!=\!
\frac{4 m_{\rm r}^3 \alpha^2 R_{\rm p}^2 \mathcal{R}}
{3 m_{\rm e}^3 \mathchar'26\mkern-9mu \lambda_C^2 n^3}
\Big\{1\!-\!
\alpha^2 \ln\frac{\alpha R_{\rm p}}{n \mathchar'26\mkern-9mu \lambda_C} 
-1.8 \alpha^2
\Big\}\delta_{\ell 0},
\label{eq:ENS}
\end{eqnarray}
\begin{eqnarray}
&&E_{\rm RR}\!=\!
\frac{ 2 m_{\rm r}^3 \alpha^4 \mathcal{R}}
{m_{\rm p} m_{\rm e}^2 \pi^2 n^3}
\delta_{\ell 0}
\Big\{
\!-\!13.47
\!+\!\frac{2}{3}\pi\alpha\lambda^2
\!+\!\Delta^{\!\rm RR}\alpha\lambda
\Big \}, \ \ \ \ \ 
\label{eq:ERR}
\end{eqnarray}
and
\begin{eqnarray}
&&E_{\rm SEN}\!=\!
\frac{ 8 m_{\rm r}^3 \alpha^3 \mathcal{R}}
{3 m_{\rm p}^2 m_{\rm e} \pi n^3}
\Big\{
\delta_{\ell 0} \ln \! \frac{m_{\rm p}}{m_{\rm e}\alpha^2}
\!-\!\beta_{n\ell}
\Big\}.
\label{eq:ESEN}
\end{eqnarray}
Here, 
$\lambda$=$\ln (\alpha^{-2} m_{\rm e}/m_{\rm r} )$,
$\delta_{\ell 0}$ is the Kronecker delta, 
${\kappa_{\ell j}=(\ell-j)(2j+1)}$,   
${f_{nj}=[1 + \alpha^2(n - \delta)^{-2}]^{-1_{\!}/_{\!}2}}$ in Eq.~(\ref{eq:EM})
(with $\delta$ = ${j+\frac{1}{2}-[(j+\frac{1}{2})^2-\alpha^2]^{1_{\!}/_{\!}2}}$),
$\beta_{n \ell}$ are the
Bethe logarithms 
(tabulated in 
Ref.~\cite{drake1990bethe}),
\begin{eqnarray}
a_{n\ell}\!=\!
-2\Big[\ln\!\frac{2}{n}\!+\!\sum_{i=1}^{n}\frac{1}{i}\!+\!1\!-\!\frac{1}{2n}\Big] \delta_{\ell 0}
\!+\!\frac{1\!-\!\delta_{\ell 0}}{\ell(\ell\!+\!1)(2\ell\!+\!1)},\ \ \ \ \ 
\label{eq:an}
\end{eqnarray}
\begin{eqnarray}
D_{60}^{(n\ell)}\!=\!
(4 \ln 2\!-\!\frac{7}{2}) \pi \delta_{\ell 0}
\!+\!
\frac{[3\!-\!\frac{\ell(\ell\!+\!1)}{n^2}]2\pi(1\!-\!\delta_{\ell 0})}{(4\ell^2\!-\!1)(2\ell\!+\!3)},\ \ \ \ \ 
\label{eq:D60}
\end{eqnarray} 
\begin{eqnarray}
&&A_{61}^{(n \ell j)}\!=
\big[
(\sum_{i=1}^{n}\frac{4}{i})
+\frac{28}{3} \ln 2
-4 \ln n
-\frac{601}{180}
-\frac{77}{45 n^2}
\big] \delta_{\ell 0}
\nonumber \\ &&
\!+(1\!-\!\frac{1}{n^2})(\frac{2}{15}\!+\!\frac{\delta_{j \frac{1}{2}}}{3})
\delta_{\ell 1}
\!+\!\frac{8(3n^2\!-\!\ell(\ell\!+\!1))(1\!-\!\delta_{\ell 0})}
{3n^2\ell(4\ell^2\!-\!1)(\ell\!+\!1)(2\ell\!+\!3)}
,
\nonumber \\
\label{eq:A61}
\end{eqnarray}
and 
\begin{eqnarray}
&&B_{62}^{(n\ell)}\!\!=\!\!
\frac{16}{9}
\big[
1.067
\!+\!\psi(n)
\!-\!\ln n
\!-\!\frac{1}{n}
\!+\!\frac{1}{4n^2}
\big]\delta_{\ell 0}
\!+\!\frac{4 \delta_{\ell 1}}{27}\frac{n^2\!-\!1}{n^2},
\
\nonumber \\
\label{eq:B62}
\end{eqnarray}
with $\psi$ being the digamma function.
The values of 
$G_{\rm SE}^{(n \ell j)}$ of Eq.~(\ref{eq:E2SE}),
$G_{\rm VP}^{(1)(n \ell j)}$ of Eq.~(\ref{eq:E2VP}),
and
$B_{61}^{(n \ell j)}$ and $B_{60}^{(n \ell j)}$ of Eq.~(\ref{eq:E4})  
are discussed in 
Ref.~\cite{mohr2012codata}, and 
the tabulated values (along with values given in 
Refs.~\cite{jentschura2006two,jentschura2005self,jentschura2004electron,
jentschura2003radiative,kotochigova2002precise}
and simple extrapolations and interpolations)
are sufficiently precise for the current work.
The values of $D_{71}^{(n \ell j)}$ are needed only 
for the lowest-lying states, and have recently been
calculated in Ref.~\cite{yerokhin2015nuclear}.

Although many uncertainties to the fine-structure energy contributions
$E^{\rm (fs)}_{n\ell j}$
are detailed in Ref.~\cite{mohr2012codata}, 
only four uncertainties dominate and 
need to be considered
in this work.
The first comes from an uncertainty of $\pm$19.7 in the 
$\ell$=0 coefficients $B_{60}^{(n \ell j)}$ 
of Eq.~(\ref{eq:E4}), which leads to an uncertainty of 
$\delta_{\ell 0}$(2.0~kHz)/$n^3$.
The second comes from the uncertainty in $R_{\rm p}$ 
(of Eq.~(\ref{eq:crema})) in Eq.~(\ref{eq:ENS}), 
which contributes
$\delta_{\ell 0}$(1.0~kHz)/$n^3$.
The third comes from the coefficient $C_{50}$
of Eq.~(\ref{eq:E6}), which is presumed to be
$\pm$30~$\delta_{\ell 0}$, and which leads to an uncertainty of 
$\delta_{\ell 0}$(1.0~kHz)/$n^3$.
The fourth comes from 
the $\Delta^{\!\rm RR}$ coefficient of 
Eq.~(\ref{eq:ERR}), which is presumed to be 
$\pm$10, 
thus leading to an uncertainty of 
$\delta_{\ell 0}$(0.7~kHz)/$n^3$.
An additional uncertainty in Eq.~(\ref{eq:ER}),
has now been resolved by Ref.~\cite{yerokhin2015nuclear}
and does not need to be included.
These four dominant uncertainties all have the same dependence on $n$
and $\ell$ and therefore can be added in quadrature to give a combined 
uncertainty of $\delta_{\ell 0}$(2.6~kHz)/$n^3$. 
All other uncertainties listed in 
Ref.~\cite{mohr2012codata} are more than an order of magnitude
smaller. 
The values of $E^{\rm (fs)}_{n\ell j}$ for the 
lowest-lying states, along with their uncertainties 
are listed in Table~\ref{table:Efs}.
 
\begin{table}[t]
\begin{ruledtabular}
\caption{\label{table:Efs} Fine-structure energies $E^{\rm (fs)}_{n\ell j}$
for the lowest-lying states of atomic hydrogen. Uncertainties in the 
last digits are shown in parentheses. These values were
determined using the Rydberg constant 
obtained in Section~\ref{sec:gross}; 
however, using the CODATA 2014 value instead
would not change the values, since the resulting changes would be
at the level of 1 Hz or less.
}
\begin{tabular}{cccc}
&$E^{\rm (fs)}_{n\ell j}$(kHz)&&$E^{\rm (fs)}_{n\ell j}$(kHz)\\
\hline
1S$_{1_{\!}/_{\!}2}$& -35\,625\,530.5(2.6)\\
2S$_{1_{\!}/_{\!}2}$& -12\,636\,029.4(3)&&\\
2P$_{1_{\!}/_{\!}2}$& -13\,693\,861.6(0)&2P$_{3_{\!}/_{\!}2}$&  -2\,724\,820.1(0)\\
3S$_{1_{\!}/_{\!}2}$&  -4\,552\,716.0(1)&&\\
3P$_{1_{\!}/_{\!}2}$&  -4\,867\,590.3(0)&3P$_{3_{\!}/_{\!}2}$&  -1\,617\,501.0(0)\\
3D$_{3_{\!}/_{\!}2}$&  -1\,622\,832.7(0)&3D$_{5_{\!}/_{\!}2}$&     -539\,495.5(0)\\
4S$_{1_{\!}/_{\!}2}$&  -2\,091\,332.8(0)&&\\
4P$_{1_{\!}/_{\!}2}$&  -2\,224\,408.7(0)&4P$_{3_{\!}/_{\!}2}$&     -853\,278.9(0)\\
4D$_{3_{\!}/_{\!}2}$&     -855\,566.5(0)&4D$_{5_{\!}/_{\!}2}$&     -398\,533.1(0)\\
4F$_{5_{\!}/_{\!}2}$&     -399\,342.4(0)&4F$_{7_{\!}/_{\!}2}$&     -170\,827.1(0)\\
5S$_{1_{\!}/_{\!}2}$&  -1\,123\,202.3(0)&&\\
5P$_{1_{\!}/_{\!}2}$&  -1\,191\,397.0(0)&5P$_{3_{\!}/_{\!}2}$&     -489\,379.3(0)\\
5D$_{3_{\!}/_{\!}2}$&     -490\,561.2(0)&5D$_{5_{\!}/_{\!}2}$&     -256\,560.1(0)\\
5F$_{5_{\!}/_{\!}2}$&     -256\,977.8(0)&5F$_{7_{\!}/_{\!}2}$&     -139\,977.9(0)\\
5G$_{7_{\!}/_{\!}2}$&     -140\,196.4(0)&5G$_{9_{\!}/_{\!}2}$&      -69\,996.6(0)\\
6S$_{1_{\!}/_{\!}2}$&     -670\,236.8(0)&&\\
6P$_{1_{\!}/_{\!}2}$&     -709\,720.8(0)&6P$_{3_{\!}/_{\!}2}$&     -303\,461.0(0)\\
6D$_{3_{\!}/_{\!}2}$&     -304\,148.6(0)&6D$_{5_{\!}/_{\!}2}$&     -168\,731.3(0)\\
6F$_{5_{\!}/_{\!}2}$&     -168\,974.3(0)&6F$_{7_{\!}/_{\!}2}$&     -101\,266.0(0)\\
6G$_{7_{\!}/_{\!}2}$&     -101\,393.0(0)&6G$_{9_{\!}/_{\!}2}$&      -60\,768.1(0)\\
6H$_{9_{\!}/_{\!}2}$&      -60\,846.8(0)&6H$_{11_{\!}/_{\!}2}$&      -33\,763.6(0)
\end{tabular}
\end{ruledtabular}
\end{table}

\section{hyperfine strurcture}
\label{sec:hfs}

The hyperfine contributions to the energies are given by
\begin{eqnarray}
E^{\rm (hfs)}_{n\ell jf}\!=\!
\delta_{\ell 0}(f\!-\!\frac{3}{4})\frac{\Delta E^{\rm hfs}_{1S}\!+\!\Delta_n}{n^3}
\!+\!(1\!-\!\delta_{\ell 0})(E_{\rm diag}^{\rm hfs}\!+\!\Delta E^{\rm hfs}).
\nonumber \\
\label{eq:hfs}
\end{eqnarray}
For ${\ell=0}$ states, 
where the structure of the nucleus affects the 
hyperfine structure,
$E^{\rm (hfs)}_{n\ell jf}$ is 
determined using the precise measurement of 
the ground-state hyperfine interval $\Delta E^{\rm hfs}_{1S}$ 
of Eq.~(\ref{eq:Hexpt}). 
For $n$=2,
$\Delta_2=48.922(27)$~kHz 
can be determined directly from the measured interval
of Eq.~(\ref{eq:Hexpt}). For higher $n$,
the correction 
$\Delta_n$
is discussed in detail in 
Ref.~\cite{jentschura2006quantum}, 
and, to the accuracy required here, is given by 
\begin{eqnarray}
\Delta_n\!=\frac{8}{3}g_{\rm p}
\alpha^4 \mathcal{R}
\frac{m_{\rm e}}{m_{\rm p}}
\Big(
\frac{1}{3}
\!+\!\frac{3}{2n}
\!-\!\frac{11}{6n^2}
\Big).
\label{eq:Deltan}
\end{eqnarray} 

For ${\ell\ne0}$, 
nuclear effects are not significant, 
and the dominant diagonal term $E_{\rm diag}^{\rm hfs}$
is given by 
\begin{eqnarray}
&&E_{\rm diag}^{\rm hfs}\!=\!g_{\rm p}
\frac{\alpha^2 \mathcal{R}}{n^3}
\frac{m_{\rm r}^3}{m_{\rm e}^3}
\frac{m_{\rm e}}{m_{\rm p}}
\frac{f(f\!+\!1)\!-\!j(j\!+\!1)\!-\!\frac{3}{4}}{(2\ell\!+\!1)j(j\!+\!1)}
\bigg\{1
\!+\!\frac{a_{\rm e}}{2\kappa_{\ell j}}
\nonumber \\
&&\ \ \ \ \ \ \ 
\!+\frac{g_{\rm p}\!-\!1}{g_{\rm p}}
\frac{m_{\rm e}}{m_{\rm p}}
\frac{2\kappa_{\ell j}\!-\!1}{2\kappa_{\ell j}}
\!+\!\alpha^2 \Big[
\frac{3(2j\!+\!1)^2\!-\!1}{2(2j\!+\!1)^2j(j\!+\!1)}
\nonumber \\
&&\ \ \ \ \ \ \ \ \ \ \ \ \ \ \ \!+\frac{3}{n(2j\!+\!1)}
\!+\!\frac{3\!-\!8\kappa_{\ell j}}{2n^2(2\kappa_{\ell j}\!-\!1)}
\Big]
\bigg\}.
\label{eq:diag}
\end{eqnarray} 

The $a_e$=${(g_e\!-\!2)/2}$ electron anomalous moment 
corrections do not apply to the 
${\vec{I} \cdot \vec{L}}$ term in the hyperfine Hamiltonian,
and this leads to the 
$2 \kappa_{\ell j}$ denominator of the $a_e$ term.
The $a_e$ correction is included 
(for the 2P states) 
in Ref.~\cite{brodsky1967precise}, 
is included to first order in $\alpha$
in Ref.~\cite{wundt2011self}, 
but is included incorrectly 
(without the ${2\kappa_{\ell j}}$ in the denominator) 
in Ref.~\cite{kramida2010critical}
and Ref.~\cite{weitz1995precision}.
Given the size of the hyperfine structure, it is sufficient  for this work to 
approximate $a_e$ by its lowest-order term: $\alpha/(2\pi)$.

The correction proportional to 
${((g_{\rm p}\!-\!1)/g_{\rm p})(m_{\rm e}/m_{\rm p})}$  
results from the interaction of the proton's orbital motion with its spin.
This term is
included (for the 2P states) in Ref.~\cite{brodsky1967precise}, 
but is given  incorrectly for the 2P state in Eqs.~(27)
and (28) of Ref.~\cite{martynenko2008fine} for muonic hydrogen,
where the term is even more important. 
The term is correctly included for muonic hydrogen 
in Table II of  Ref.~\cite{martynenko2008fine}
and in Ref.~\cite{jentschura2011lamb}. 
The term is omitted in 
Ref.~\cite{wundt2011self}, Ref.~\cite{wundt2010proposal}, Ref~\cite{weitz1995precision} and Ref.~\cite{kramida2010critical}.  
This mass-correction term contributes 13 kHz 
to the 2P$_{1_{\!}/_{\!}2}$ hyperfine structure, and thus must certainly
be included at the accuracy of this work. 
The general form for this correction (as a function of $n$, $\ell$, and
$j$) does not seem to appear
previously in the literature. 

The correction proportional to $\alpha^2$  
in Eq.~(\ref{eq:diag}) 
is a relativistic correction 
which is given in Ref.~\cite{wundt2011self}. 
Higher-order corrections
are also considered in that work, but are insignificant at the level of this
work.

\begin{table}[t]
\begin{ruledtabular}
\caption{\label{table:Ehfs} Hyperfine-structure energies $E^{\rm (hfs)}_{n\ell j f}$
for the lowest-lying states of atomic hydrogen. 
All values are uncertain by less than 0.1 kHz.
The precisely measured values are given for the 1S and 2S states.
}
\begin{tabular}{ccccc}
&$f$&$E^{\rm (hfs)}_{n\ell j f}$(kHz)&$f$&$E^{\rm (hfs)}_{n\ell j f}$(kHz)\\
\hline
1S$_{1_{\!}/_{\!}2}$&0&$-1\,065\,304.313\,8260(8)$&
1&$355\,101.437\,9420(3)$\\
2S$_{1_{\!}/_{\!}2}$&0&$-133\,167.6257(51)$&
1&$44\,389.2086(17)$\\
2P$_{1_{\!}/_{\!}2}$&0&$-44\,379.0$&
1&$14\,790.5$\\
2P$_{3_{\!}/_{\!}2}$&1&$-14\,781.3$&
2&$8\,870.3$\\
3S$_{1_{\!}/_{\!}2}$&0&$-39\,457.0$&
1&$13\,152.3$\\
3P$_{1_{\!}/_{\!}2}$&0&$-13\,149.4$&
1&$4\,382.4$\\
3P$_{3_{\!}/_{\!}2}$&1&$-4\,379.7$&
2&$2\,628.2$\\
3D$_{3_{\!}/_{\!}2}$&1&$-2\,629.2$&
2&$1\,577.4$\\
3D$_{5_{\!}/_{\!}2}$&2&$-1\,576.9$&
3&$1\,126.4$\\
4S$_{1_{\!}/_{\!}2}$&0&$-16\,645.9$&
1&$5\,548.6$\\
4P$_{1_{\!}/_{\!}2}$&0&$-5\,547.4$&
1&$1\,848.8$\\
4P$_{3_{\!}/_{\!}2}$&1&$-1\,847.7$&
2&$1\,108.8$\\
4D$_{3_{\!}/_{\!}2}$&1&$-1\,109.2$&
2&$665.5$\\
4D$_{5_{\!}/_{\!}2}$&2&$-665.3$&
3&$475.2$\\
4F$_{5_{\!}/_{\!}2}$&2&$-475.3$&
3&$339.5$\\
4F$_{7_{\!}/_{\!}2}$&3&$-339.4$&
4&$264.0$
\end{tabular}
\end{ruledtabular}
\end{table}

The smaller $\Delta E^{\rm hfs}$ contribution comes from
an off-diagonal 
element of the hyperfine Hamiltonian. 
This element causes a very slight mixing between the 
${|n \ \ell \  j\!=\!\ell\!-\!\frac{1}{2}\  f\!=\!\ell \ m_f \rangle}$ 
state and the 
${|n \ \ell \  j\!=\!\ell\!+\!\frac{1}{2}\  f\!=\!\ell \ m_f \rangle}$ 
state, and its contribution to the energy can be determined to 
sufficient accuracy
by the expression from second-order perturbation theory:
\begin{eqnarray}
\Delta E^{\rm hfs}&&=
\frac{|\langle n \ \ell \  j \  f\!=\!\ell \  m_f
 |H_{\rm hfs} | n \ \ell \  j'\  f\!=\!\ell \ m_f \rangle
|^2}{E^{\rm (fs)}_{n\ell j}-E^{\rm (fs)}_{n \ell j'} }
\nonumber \\
&&=
\frac{2\alpha^2 \mathcal{R}}{n^3}
\frac{m_{\rm e}^2}{m_{\rm p}^2}
g_{\rm p}^2
\frac{(j\!-\!\ell)\delta_{f \ell}}{(2\ell\!+\!1)^4}.
\label{eq:DeltaEhfs}
\end{eqnarray}
This expression for $\Delta E^{\rm hfs}$ was first 
given (for the 2P states) in Ref.~\cite{brodsky1967precise}.
Its general form does not appear to be presented in the literature,
and an incorrect form 
(with an incorrect dependence on $\ell$)  
is used in the tabulation of Ref.~\cite{kramida2010critical}. 
The contribution used in Ref.~\cite{kramida2010critical} 
for the off-diagonal contributions is too large by a  
factor of $\ell^2 (\ell+1)^2 (2\ell+1)^2/36$ (a 
factor of 25, 196, 900, and 3025 for 
D, F, G, and H states, respectively). 

The values of $E^{\rm (hfs)}_{n\ell jf}$ of Eq.~(\ref{eq:hfs}) are listed 
in Table~\ref{table:Ehfs}. The tabulated values 
include the contributions Eq.~(\ref{eq:DeltaEhfs}). 
In all cases the values can be determined to better than
0.1 kHz.

\section{gross structure}
\label{sec:gross}

The gross structure $E^{\rm (g)}$ 
of Eq.~(\ref{eq:Eg}) requires a precise value for the 
Rydberg constant. 
This value can be obtained by equating the precise measured value 
of Eq.~(\ref{eq:Hexpt})
for the 
${1{\rm S}_{1_{\!}/_{\!}2} f\!=\!1\!\to\!2{\rm S}_{1_{\!}/_{\!}2} f\!=\!1}$ 
interval to Eq.~(\ref{eq:Henergy}), with the values 
of $E_{n \ell j}^{\rm (fs)}$ and $E_{n \ell j f}^{\rm (hfs)}$ obtained in 
the previous sections:
\begin{eqnarray}
&&2\,466\,061\,102\,474.806(10)\ {\rm kHz}=
\frac{3}{4}
\mathcal{R}
\frac{m_{\rm r}}{m_{\rm e}} 
\nonumber \\
&&
+22\,989\,501.2(2.2)\  {\rm kHz}
-310\,712.2294(17)\  {\rm kHz}.
\label{eq:determineRy}
\end{eqnarray}  
The first number in the second line of Eq.~(\ref{eq:determineRy})
is due to the difference of $E_{n \ell j}^{\rm (fs)}$ for the two
states, and includes the correlated error for the difference. The second
term is due to the difference of $E_{n \ell j f}^{\rm (hfs)}$ for the two
states, and for these 1S and 2S states, $E^{\rm (hfs)}$ is known
precisely from the experimental results of Eq.~(\ref{eq:Hexpt}).
 Solving for the Rydberg constant gives:
\begin{eqnarray}
\mathcal{R} = 
c R_{\infty} = 3\,289\,841\,960\,248.9(3.0) \  {\rm kHz},
\label{eq:Ry}
\end{eqnarray}
where the uncertainty is dominated by the uncertainty in 
the $E^{\rm (fs)}$ term.
Eq.~(\ref{eq:Ry}) differs considerably from the CODATA 2014 value of 
${3\,289\,841\,960\,355(19)}\  {\rm kHz}$. The difference is almost
entirely due to the fact that the CREMA value of $R_{\rm p}$
(Eq.~(\ref{eq:crema})) is used, whereas the CODATA compilation
uses the value of Eq.~(\ref{eq:RpCODATA}). A very small
contribution to the difference comes from the recent improvement
\cite{yerokhin2015nuclear}
in the determination of $E^{\rm (fs)}$ from the determination 
of $D_{71}^{(n \ell j)}$.
The uncertainty in the CREMA value of $R_{\rm p}$ 
(Eq.~(\ref{eq:crema})) contributes 1.2~kHz to the 3.0~kHz
uncertainty in Eq.~(\ref{eq:Ry}), with the rest of the uncertainty 
resulting from the other (theoretical) uncertainties discussed 
in the last paragraph of Section~\ref{sec:fs}.

\section{total binding energies}
\label{sec:total}

\begin{table}[t]
\begin{ruledtabular}
\caption{\label{table:ES} Total binding energies for the lowest-lying S
($\ell=0$) 
states, 
with uncertainties in the last digits given in parentheses. 
The last column gives the change $\delta_R$ in the binding energy that would 
result if $R_{\rm p}$ is increased by 0.03423 fm (the difference between
the CODATA and CREMA values).
}
\begin{tabular}{rrrc}
$n$&
$E$($n$S$_{1_{\!}/_{\!}2}f\!=\!0)$ (kHz)&
$E$($n$S$_{1_{\!}/_{\!}2}f\!=\!1)$ (kHz)&
$\delta_R$(kHz)\\
\hline
1&  -3\,288\,087\,922\,416.0(4)&  -3\,288\,086\,502\,010.2(4)&                     -15.3\\
2&     -822\,025\,577\,092.2(4)&     -822\,025\,399\,535.4(4)&                     -15.3\\
3&     -365\,343\,617\,904.3(2)&     -365\,343\,565\,294.9(2)&                      -8.5\\
4&     -205\,505\,309\,952.5(1)&     -205\,505\,287\,757.9(1)&                      -5.3\\
5&     -131\,523\,180\,988.2(1)&     -131\,523\,169\,624.6(1)&                      -3.6\\
6&      -91\,335\,431\,601.7(1)&      -91\,335\,425\,025.6(1)&                      -2.6\\
7&      -67\,103\,520\,641.6(1)&      -67\,103\,516\,500.4(1)&                      -1.9\\
8&      -51\,376\,096\,003.6(0)&      -51\,376\,093\,229.3(0)&                      -1.5\\
9&      -40\,593\,435\,126.6(0)&      -40\,593\,433\,178.1(0)&                      -1.2\\
10&      -32\,880\,666\,896.5(0)&      -32\,880\,665\,476.0(0)&                      -1.0\\
11&      -27\,174\,094\,155.3(0)&      -27\,174\,093\,088.1(0)&                      -0.8\\
12&      -22\,833\,779\,830.0(0)&      -22\,833\,779\,008.0(0)&                      -0.7\\
13&      -19\,455\,996\,185.2(0)&      -19\,455\,995\,538.7(0)&                      -0.6\\
14&      -16\,775\,829\,289.9(0)&      -16\,775\,828\,772.2(0)&                      -0.5\\
15&      -14\,613\,608\,126.8(0)&      -14\,613\,607\,705.9(0)&                      -0.4\\
16&      -12\,843\,989\,064.0(0)&      -12\,843\,988\,717.2(0)&                      -0.4\\
17&      -11\,377\,372\,464.9(0)&      -11\,377\,372\,175.8(0)&                      -0.4\\
18&      -10\,148\,333\,775.8(0)&      -10\,148\,333\,532.3(0)&                      -0.3\\
19&       -9\,108\,198\,599.8(0)&       -9\,108\,198\,392.7(0)&                      -0.3\\
20&       -8\,220\,148\,221.6(0)&       -8\,220\,148\,044.1(0)&                      -0.3
\end{tabular}
\end{ruledtabular}
\end{table}

\begin{table*}[t]
\begin{ruledtabular}
\caption{\label{table:EP} Continuation of 
Table~\ref{table:ES} for 
the lowest-lying P  ($\ell=1$) states, 
}
\begin{tabular}{rrrrrc}
$n$&
$E$($n$P$_{1_{\!}/_{\!}2} f\!=\!0)$ (kHz)&
$E$($n$P$_{1_{\!}/_{\!}2} f\!=\!1)$ (kHz)&
$E$($n$P$_{3_{\!}/_{\!}2} f\!=\!1)$ (kHz)&
$E$($n$P$_{3_{\!}/_{\!}2} f\!=\!2)$ (kHz)&
$\delta_R$(kHz)\\
\hline
2&-822\,026\,546\,135.9(7)&-822\,026\,486\,966.4(7)&-822\,015\,547\,496.7(7)&-822\,015\,523\,845.1(7)&-26.8\\
3&-365\,343\,906\,471.0(3)&-365\,343\,888\,939.2(3)&-365\,340\,647\,611.9(3)&-365\,340\,640\,604.0(3)&-11.9\\
4&-205\,505\,431\,929.9(2)&-205\,505\,424\,533.7(2)&-205\,504\,057\,100.4(2)&-205\,504\,054\,143.9(2)& -6.7\\
5&-131\,523\,243\,500.5(1)&-131\,523\,239\,713.6(1)&-131\,522\,539\,588.6(1)&-131\,522\,538\,074.9(1)& -4.3\\
6& -91\,335\,467\,797.2(1)& -91\,335\,465\,605.8(1)& -91\,335\,060\,441.3(1)& -91\,335\,059\,565.3(1)& -3.0\\
7& -67\,103\,543\,442.8(1)& -67\,103\,542\,062.8(1)& -67\,103\,286\,915.7(1)& -67\,103\,286\,364.0(1)& -2.2\\
8& -51\,376\,111\,281.9(0)& -51\,376\,110\,357.4(0)& -51\,375\,939\,428.9(0)& -51\,375\,939\,059.3(0)& -1.7\\
9& -40\,593\,445\,858.7(0)& -40\,593\,445\,209.3(0)& -40\,593\,325\,160.9(0)& -40\,593\,324\,901.3(0)& -1.3\\
10& -32\,880\,674\,721.0(0)& -32\,880\,674\,247.6(0)& -32\,880\,586\,732.3(0)& -32\,880\,586\,543.1(0)& -1.1\\
11& -27\,174\,100\,034.5(0)& -27\,174\,099\,678.8(0)& -27\,174\,033\,927.3(0)& -27\,174\,033\,785.2(0)& -0.9\\
12& -22\,833\,784\,358.7(0)& -22\,833\,784\,084.8(0)& -22\,833\,733\,439.4(0)& -22\,833\,733\,329.9(0)& -0.7\\
13& -19\,455\,999\,747.3(0)& -19\,455\,999\,531.9(0)& -19\,455\,959\,697.9(0)& -19\,455\,959\,611.8(0)& -0.6\\
14& -16\,775\,832\,142.0(0)& -16\,775\,831\,969.5(0)& -16\,775\,800\,076.2(0)& -16\,775\,800\,007.3(0)& -0.5\\
15& -14\,613\,610\,445.7(0)& -14\,613\,610\,305.5(0)& -14\,613\,584\,375.1(0)& -14\,613\,584\,319.0(0)& -0.5
\end{tabular}
\end{ruledtabular}
\end{table*}

\begin{table*}
\begin{ruledtabular}
\caption{\label{table:ED} Continuation of 
Tables~\ref{table:ES}~and~\ref{table:EP} for
the lowest-lying D  ($\ell=2$) states.
}
\begin{tabular}{rrrrrc}
$n$&
$E$($n$D$_{3_{\!}/_{\!}2} f\!=\!1)$ (kHz)&
$E$($n$D$_{3_{\!}/_{\!}2} f\!=\!2)$ (kHz)&
$E$($n$D$_{5_{\!}/_{\!}2} f\!=\!2)$ (kHz)&
$E$($n$D$_{5_{\!}/_{\!}2} f\!=\!3)$ (kHz)&
$\delta_R$(kHz)\\
\hline
3&-365\,340\,651\,193.1(3)&-365\,340\,646\,986.5(3)&-365\,339\,566\,803.7(3)&-365\,339\,564\,100.3(3)&-11.9\\
4&-205\,504\,058\,649.5(2)&-205\,504\,056\,874.8(2)&-205\,503\,601\,172.2(2)&-205\,503\,600\,031.7(2)& -6.7\\
5&-131\,522\,540\,392.4(1)&-131\,522\,539\,483.7(1)&-131\,522\,306\,163.9(1)&-131\,522\,305\,580.0(1)& -4.3\\
6& -91\,335\,060\,910.1(1)& -91\,335\,060\,384.3(1)& -91\,334\,925\,361.2(1)& -91\,334\,925\,023.3(1)& -3.0\\
7& -67\,103\,287\,212.4(1)& -67\,103\,286\,881.3(1)& -67\,103\,201\,852.2(1)& -67\,103\,201\,639.4(1)& -2.2\\
8& -51\,375\,939\,628.3(0)& -51\,375\,939\,406.5(0)& -51\,375\,882\,443.7(0)& -51\,375\,882\,301.1(0)& -1.7\\
9& -40\,593\,325\,301.3(0)& -40\,593\,325\,145.5(0)& -40\,593\,285\,138.7(0)& -40\,593\,285\,038.6(0)& -1.3\\
10& -32\,880\,586\,834.9(0)& -32\,880\,586\,721.3(0)& -32\,880\,557\,556.4(0)& -32\,880\,557\,483.4(0)& -1.1\\
11& -27\,174\,034\,004.5(0)& -27\,174\,033\,919.1(0)& -27\,174\,012\,007.1(0)& -27\,174\,011\,952.2(0)& -0.9\\
12& -22\,833\,733\,498.9(0)& -22\,833\,733\,433.2(0)& -22\,833\,716\,555.3(0)& -22\,833\,716\,513.1(0)& -0.7
\end{tabular}
\end{ruledtabular}
\end{table*}
\begin{table*}[!htb]
\begin{ruledtabular}
\caption{\label{table:EhL} Continuation of 
Tables~\ref{table:ES},~\ref{table:EP},~and~\ref{table:ED} for 
the lowest-lying ($\ell \ge 3$) states. 
}
\begin{tabular}{rrrrrrc}
$n$&
$\ell$&
$E$($n\ \ell\  j\!=\!\ell\!-\!\frac{1}{2}\  f\!=\!\ell\!-\!1)$&
$E$($n\ \ell\  j\!=\!\ell\!-\!\frac{1}{2}\  f\!=\!\ell)$&
$E$($n\ \ell\  j\!=\!\ell\!+\!\frac{1}{2}\  f\!=\!\ell)$&
$E$($n\ \ell\  j\!=\!\ell\!+\!\frac{1}{2}\  f\!=\!\ell\!+\!1)$&
$\delta_R$(kHz)\\
\hline
4&3&-205\,503\,601\,791.5(2)&-205\,503\,600\,976.7(2)&-205\,503\,373\,140.3(2)&-205\,503\,372\,536.9(2)& -6.7\\
5&3&-131\,522\,306\,484.4(1)&-131\,522\,306\,067.2(1)&-131\,522\,189\,415.0(1)&-131\,522\,189\,106.0(1)& -4.3\\
6&3& -91\,334\,925\,548.0(1)& -91\,334\,925\,306.5(1)& -91\,334\,857\,799.4(1)& -91\,334\,857\,620.6(1)& -3.0\\
7&3& -67\,103\,201\,970.3(1)& -67\,103\,201\,818.3(1)& -67\,103\,159\,306.5(1)& -67\,103\,159\,193.9(1)& -2.2\\
8&3& -51\,375\,882\,523.1(0)& -51\,375\,882\,421.2(0)& -51\,375\,853\,941.6(0)& -51\,375\,853\,866.2(0)& -1.7\\
9&3& -40\,593\,285\,194.6(0)& -40\,593\,285\,123.0(0)& -40\,593\,265\,120.9(0)& -40\,593\,265\,067.9(0)& -1.3\\
5&4&-131\,522\,189\,594.9(1)&-131\,522\,189\,354.5(1)&-131\,522\,119\,365.0(1)&-131\,522\,119\,173.9(1)& -4.3\\
6&4& -91\,334\,857\,904.1(1)& -91\,334\,857\,765.0(1)& -91\,334\,817\,261.8(1)& -91\,334\,817\,151.2(1)& -3.0\\
7&4& -67\,103\,159\,372.6(1)& -67\,103\,159\,285.1(1)& -67\,103\,133\,778.7(1)& -67\,103\,133\,709.0(1)& -2.2\\
8&4& -51\,375\,853\,986.1(0)& -51\,375\,853\,927.4(0)& -51\,375\,836\,840.1(0)& -51\,375\,836\,793.4(0)& -1.7\\
9&4& -40\,593\,265\,152.2(0)& -40\,593\,265\,111.0(0)& -40\,593\,253\,110.0(0)& -40\,593\,253\,077.2(0)& -1.3\\
6&5& -91\,334\,817\,329.4(1)& -91\,334\,817\,238.9(1)& -91\,334\,790\,237.1(1)& -91\,334\,790\,161.9(1)& -3.0\\
7&5& -67\,103\,133\,821.4(1)& -67\,103\,133\,764.4(1)& -67\,103\,116\,760.4(1)& -67\,103\,116\,713.0(1)& -2.2\\
8&5& -51\,375\,836\,868.8(0)& -51\,375\,836\,830.6(0)& -51\,375\,825\,439.2(0)& -51\,375\,825\,407.5(0)& -1.7\\
9&5& -40\,593\,253\,130.2(0)& -40\,593\,253\,103.4(0)& -40\,593\,245\,102.8(0)& -40\,593\,245\,080.6(0)& -1.3\\
7&6& -67\,103\,116\,790.3(1)& -67\,103\,116\,750.2(1)& -67\,103\,104\,604.6(1)& -67\,103\,104\,570.3(1)& -2.2\\
8&6& -51\,375\,825\,459.3(0)& -51\,375\,825\,432.5(0)& -51\,375\,817\,295.9(0)& -51\,375\,817\,272.9(0)& -1.7\\
9&6& -40\,593\,245\,117.0(0)& -40\,593\,245\,098.1(0)& -40\,593\,239\,383.5(0)& -40\,593\,239\,367.4(0)& -1.3\\
8&7& -51\,375\,817\,310.8(0)& -51\,375\,817\,290.9(0)& -51\,375\,811\,188.4(0)& -51\,375\,811\,171.0(0)& -1.7\\
9&7& -40\,593\,239\,394.0(0)& -40\,593\,239\,380.0(0)& -40\,593\,235\,094.1(0)& -40\,593\,235\,081.9(0)& -1.3\\
9&8& -40\,593\,235\,102.2(0)& -40\,593\,235\,091.4(0)& -40\,593\,231\,757.9(0)& -40\,593\,231\,748.3(0)& -1.3
\end{tabular}
\end{ruledtabular}
\end{table*}

Using Eq.~(\ref{eq:Henergy}), along with the Rydberg constant
of Eq.~(\ref{eq:Ry}), the values of $E_{n \ell j}^{\rm (fs)}$ of 
Section~\ref{sec:fs}
and $E_{n \ell j f}^{\rm (hfs)}$ of Section~\ref{sec:hfs},
allows for a determination of the energies of all bound states 
of atomic hydrogen. Energies for $\ell$=0, 1, and 2 
are given in Tables~\ref{table:ES}, \ref{table:EP},
and \ref{table:ED}, respectively, with higher-$\ell$
energies given in Table~\ref{table:EhL}. The uncertainties
listed are dominated by the uncertainties in $E^{\rm (fs)}$
(both due to the fine structure of the state and due to the 
effect of $E^{\rm (fs)}$ on the determination of the Rydberg
constant). The uncertainties listed take into account the correlation
between these two ways that $E^{\rm (fs)}$ enters the 
determination of the energies. 

The final column in the tables
gives the shift that the energy levels would experience if 
the CODATA 2014 value of $R_{\rm p}$ were used instead 
of the CREMA value. In referring to Eq.~(\ref{eq:ENS}), 
one might assume that the value of 
$R_{\rm p}$ would affect only $\ell$=0 states. 
However, this is not the case, since the value of $R_{\rm p}$
also affects the determination of the Rydberg constant 
(see Eq.~(\ref{eq:determineRy})), and therefore, 
the energies of all states
are affected. 

\section{comparison to measured values}

Table~\ref{table:Compare} gives a compilation of the most
precise measurements to date of atomic hydrogen intervals. 
Many of these measurements reported values that were
corrected for hyperfine structure. Given the inconsistency
of hyperfine corrections applied in the literature
(including inconsistent or incorrect application of 
anomalous moment corrections, of finite mass corrections 
and of corrections due to off-diagonal hyperfine-structure 
contributions), we have,
where possible, given the actual intervals 
(or linear combination of intervals) measured.

\begin{table*}[!htp]
\begin{ruledtabular}
\caption{\label{table:Compare}  Comparison to Measurements. 
Column 2 gives the measured interval (or linear combination of 
intervals), and column 3 gives the predicted interval from this work.
The final column gives the amount by which the proton radius 
would have to change in order to give agreement between
column 2 and 3. One standard deviation uncertainties are given in 
parentheses.
}
\begin{tabular}{lrrr}
Interval (or combinations of intervals)&
Measurement (kHz)&
This Work (kHz)&
$\Delta R_{\rm p}$(fm)\\
\hline
(2S$_{\!\frac{1}{2}}f$=1$\to$4S$_{\!\frac{1}{2}}f
$=1)$-\frac{1}{4}$(1S$_{\!\frac{1}{2}}f$=1$\to$2S$_{\!\frac{1}{2}}f$=1)& $        4\,836\,176( 10) $\footnotemark[1] & $        4\,836\,158.8(3) $ & $+0.059(34)$ \\
$\frac{2}{9}$(2S$_{\!\frac{1}{2}}f$=1$\to$4D$_{\!\frac{5}{2}}f$=2)$
+\!\frac{7}{9}$(2S$_{\!\frac{1}{2}}f$=1$\to$4D$_{\!\frac{5}{2}}f$=3)$-\frac
{1}{4}$(1S$_{\!\frac{1}{2}}f$=1$\to$2S$_{\!\frac{1}{2}}f$=1)\!\!\!\!\!\!\!\!\!\!& $        6\,523\,655( 24) $\footnotemark[2] & $        6\,523\,631.6(2) $ & $+0.093(95)$ \\
2S$_\frac{1}{2}f$=1$\to$8S$_\frac{1}{2}f$=1& $ 770\,649\,306\,316.4(  8.6) $\footnotemark[3] & $ 770\,649\,306\,306.1(4) $ & $+0.025(21)$ \\
$\frac{3}{8}$(2S$_\frac{1}{2}f$=1$\to$8D$_\frac{3}{2}
f$=1)+$\frac{5}{8}$(2S$_\frac{1}{2}f$=1$\to$8D$_\frac{3}{2}f$=2)& $ 770\,649\,460\,060.8(  8.3)$\footnotemark[4] & $ 770\,649\,460\,045.7(4) $ & $+0.038(21)$ \\
$\frac{5}{12}$(2S$_\frac{1}{2}f$=1$\to$8D$_\frac{5}{2}
f$=2)+$\frac{7}{12}$(2S$_\frac{1}{2}f$=1$\to$8D$_\frac{5}{2}f$=3)& $ 770\,649\,517\,195.0(  6.4)$\footnotemark[5] & $ 770\,649\,517\,174.9(4) $ & $+0.051(16)$ \\
$\frac{3}{8}$(2S$_\frac{1}{2}f$=1$\to$12D$_\frac{3}{2}
f$=1)+$\frac{5}{8}$(2S$_\frac{1}{2}f$=1$\to$12D$_\frac{3}{2}f$=2)& $ 799\,191\,666\,083.5(  9.3)$\footnotemark[6] & $ 799\,191\,666\,077.6(4) $ & $+0.014(22)$ \\
$\frac{5}{12}$(2S$_\frac{1}{2}f$=1$\to$12D$_\frac{5}{2}
f$=2)+$\frac{7}{12}$(2S$_\frac{1}{2}f$=1$\to$12D$_\frac{5}{2}f$=3)& $ 799\,191\,683\,014.5(  7.0)$\footnotemark[7] & $ 799\,191\,683\,004.7(4) $ & $+0.023(16)$ \\
1S$_\frac{1}{2}f$=1$\to$3S$_\frac{1}{2}f$=1& $ 2\,922\,742\,936\,729( 13) $\footnotemark[8] & $ 2\,922\,742\,936\,715.0(2) $ & $+0.069(65)$ \\
(2S$_\frac{1}{2}f$=1$\to$6S$_\frac{1}{2}f$=1)
$\!-\!\frac{1}{4}$(1S$_\frac{1}{2}f$=1$\to$3S$_\frac{1}{2}f$=1)& $       
 4\,240\,346(21)$\footnotemark[9] & $        4\,240\,331.0(3) $ & $+0.047(65)$ \\
$\frac{5}{12}$(2S$_{\!\frac{1}{2}}f$=1$\to$6D$_{\!\frac{5}{2}}f$=2)$
+\!\frac{7}{12}$(2S$_{\!\frac{1}{2}}f$=1$\to$6D$_{\!\frac{5}{2}}f$=3)$-\frac
{1}{4}$(1S$_{\!\frac{1}{2}}f$=1$\to$3S$_{\!\frac{1}{2}}f$=1)\!\!\!\!\!\!\!\!\!\!&
$           4\,740\,197(11)$\footnotemark[10] & $        4\,740\,192.5(3) $ & $+0.015(35)$ \\
$\frac{1}{4}$(2S$_{\!\frac{1}{2}}f$=1$\to$4P$_{\!\frac{1}{2}}
f$=0)+$\frac{3}{4}$(2S$_{\!\frac{1}{2}}f$=1$\to$4P$_{\!\frac{1}{2}}
f$=1)$-\frac{1}{4}$(1S$_{\!\frac{1}{2}}f$=1$\to$2S$_{\!\frac{1}{2}}f$=1)
\!\!\!\!\!\!\!\!\!\!& $        4\,697\,560.0( 14.9)$\footnotemark[11] & $        4\,697\,534.0(2) $ & $+0.104(59)$ \\
$\frac{3}{8}$(2S$_{\!\frac{1}{2}}f$=1$\to$4P$_{\!\frac{3}{2}}
f$=1)+$\frac{5}{8}$(2S$_{\!\frac{1}{2}}f$=1$\to$4P$_{\!\frac{3}{2}}
f$=2)$-\frac{1}{4}$(1S$_{\!\frac{1}{2}}f$=1$\to$2S$_{\!\frac{1}{2}}f$=1)
\!\!\!\!\!\!\!\!\!\!& $        6\,068\,664.0( 10.5)$\footnotemark[12] & $        6\,068\,664.1(2) $ & $-0.001(42)$ \\
2S$_{\!\frac{1}{2}}f$=0$\to$2P$_{\!\frac{3}{2}}f$=1& $       10\,029\,586( 12)$\footnotemark[13] & $       10\,029\,595.6(3) $ & $+0.029(36)$ \\
2P$_{\!\frac{1}{2}}f$=1$\to$2S$_{\!\frac{1}{2}}f$=0& $           909\,887(  9) $\footnotemark[14] & $           909\,874.2(3) $ & $+0.038(27)$ \\
2P$_{\!\frac{1}{2}}f$=1$\to$2S$_{\!\frac{1}{2}}f$=0& $           909\,904( 20)$\footnotemark[15] & $           909\,874.2(3) $ & $+0.089(60)$ 
\end{tabular}
\end{ruledtabular}
\footnotetext[1]{Ref. \cite{weitz1995precision} 
with their hfs correction of $-38\,838$ kHz  
subtracted out to give the original measured value.}
\footnotetext[2]{Refs. \cite{weitz1995precision,weitz1993solid} 
with their hfs correction of $-33.511$ kHz  
subtracted out. 
The 4D$_{5_{\!}/_{\!}2}$ hfs is not resolved in the measurement.}
\footnotetext[3]{Refs. \cite{de1997absolute,de2000metrology}
with their hfs correction of $43\,695.6$ kHz
subtracted out to give the original measured value.}
\footnotetext[4]{Refs. \cite{de1997absolute,de2000metrology}
with their hfs correction of $44\,389.2$ kHz subtracted out.
The 8D$_{3_{\!}/_{\!}2}$ hfs is not resolved in the measurement.}
\footnotetext[5]{Refs. \cite{de1997absolute,de2000metrology}
with their hfs correction of $44\,389.2$ kHz subtracted out.
The 8D$_{5_{\!}/_{\!}2}$ hfs is not resolved in the measurement.}
\footnotetext[6]{Refs. \cite{schwob1999optical,de2000metrology}
with their hfs correction of $44\,389.2$ kHz subtracted out.
The 12D$_{3_{\!}/_{\!}2}$ hfs is not resolved in the measurement.}
\footnotetext[7]{Refs. \cite{schwob1999optical,de2000metrology}
with their hfs correction of $44\,389.2$ kHz subtracted out.
The 12D$_{5_{\!}/_{\!}2}$ hfs is not resolved in the measurement.}
\footnotetext[8]{Ref. \cite{arnoult2010optical}.}
\footnotetext[9]{Ref. \cite{bourzeix1996high,de2000metrology} with their 
hfs correction of $-42\,742.1$ kHz
subtracted out to give the original measured value.}
\footnotetext[10]{Ref. \cite{bourzeix1996high,de2000metrology} with their 
hfs correction of $-41\,098.1$ kHz
subtracted out.
The 8D$_{5_{\!}/_{\!}2}$ hfs is not resolved in the measurement.}
\footnotetext[11]{Ref. \cite{berkeland1995precise} with their 
hfs correction of 
$-33\,291$ kHz subtracted out.
The 4P$_{1_{\!}/_{\!}2}$ hfs is not resolved in the measurement.}
\footnotetext[12]{Ref. \cite{berkeland1995precise} with their 
hfs correction of 
$-33\,291$ kHz subtracted out.
The 4P$_{3_{\!}/_{\!}2}$ hfs is not resolved in the measurement.}
\footnotetext[13]{Ref. \cite{hagley1994separated} 
with their hfs correction of $-118\,386$ kHz  
subtracted out to give the original measured value.}
\footnotetext[14]{Ref. \cite{lundeen1981measurement}.}
\footnotetext[15]{Ref. \cite{newton1979precision}.}
\end{table*}

Note that the hyperfine structure for the 
4P$_{1_{\!}/_{\!}2}$, 4P$_{3_{\!}/_{\!}2}$, 4D$_{5_{\!}/_{\!}2}$,
6D$_{5_{\!}/_{\!}2}$, 8D$_{3_{\!}/_{\!}2}$, 8D$_{5_{\!}/_{\!}2}$,
12D$_{3_{\!}/_{\!}2}$, and 12D$_{5_{\!}/_{\!}2}$ states
(with splittings of 
7396.2, 2956.5, 1140.5, 337.9, 221.8, 142.6, 65.7, 42.2
kHz, respectively)
is not resolved in the measurements of Table~\ref{table:Compare}, 
and therefore it is crucial 
to know both the hyperfine splittings and the relative strength 
of transitions to determine the energy splittings. 
Refs.~\cite{weitz1995precision,weitz1993solid}
give explicit values for the expected strength of the two hyperfine
transitions ($\frac{2}{9}$ and $\frac{7}{9}$) and give
an explicit correction for how much this linear combination 
differs from the hyperfine centroid of the D states.
Refs.~\cite{de2000metrology,schwob1999optical,
berkeland1995precise} do not give such explicit corrections since
their fits include the hyperfine structure. We therefore list the 
hyperfine centroid values given in those works. A further analysis
of the work of Ref.~\cite{berkeland1995precise} may be necessary,
as they appear to use a simplified form for
the presumed hyperfine intervals for their fits 
\cite{BerkelandThesis}.

The third column of Table~\ref{table:Compare} gives the 
prediction of this work for each of the measured intervals 
(or linear combination of intervals). Ten of the 15 entries 
in the table show agreement to within 1.5 standard deviations
with the values given in this work.
Four more of the entries agree to within 2 standard deviations.
One measurement, (the 2S to 12D$_{5/2}$ interval, 
which is the most precise measurement in the table)
disagrees by more than 3 standard deviations. 
The overall level of agreement
between the measured values and the values of this work can be 
assessed by calculating a $\chi^2$ value for the  
agreement for the 15 entries in the table.
The resultant $\chi^2$ of 30.6
shows that the agreement is not good. 
The $\chi^2$ distribution for 15 degrees of freedom 
has a probability  of 0.10\% for $\chi^2$ being 
30.6 or larger (which is the equivalent of a
2.3-standard-deviation effect). This 2.3-standard-deviation 
discrepancy is dominated by the 2S to 12D$_{5/2}$ interval.
If it were not included, the $\chi^2$ value would be 20.7, 
which would be a 1.2-standard-deviation discrepancy.

The 2.3-standard-deviation discrepancy between column 2 and 
column 3 of Table~\ref{table:Compare} could be eliminated if either the 
measurement of theoretical uncertainties are underestimated.
In order to make the agreement acceptable, the theoretical
uncertainty of column 3 would have to be increased by 
a factor of 20. This could be achieved by assuming an 
uncertainty (cf. the last paragraph of Section~\ref{sec:fs})
of (50 kHz)$\delta_{\ell 0}/n^3$. An overlooked contribution of
this size seems unlikely given the many decades of careful work
on atomic hydrogen theory. In order to get acceptable agreement,
the measurement uncertainties would have to be increased by 
a factor of 2. An increase in experimental uncertainties by a 
factor of 2 is far more plausible than an increase in theoretical
uncertainties by a factor of 20, especially given the large 
contribution of systematic effects in the measurements and 
given the fact that the measurement uncertainty is, in all cases,
a very small fraction of the natural width of the transition.
Ref.~\cite{karshenboim2015accuracy} discusses the tension
between the most precise measurements in Table~\ref{table:Compare}
(and includes similar measurements in deuterium), 
and suggests that the tension might indicate an
underestimation of experimental uncertainties. 

Another way to analyze the discrepancy between column 2 and 
column 3 of Table~\ref{table:Compare} is to determine the 
required change $\Delta R_{\rm p}$ in the proton radius 
(from its CREMA value) 
that would lead to agreement between the values in these columns. 
These $\Delta R_{\rm p}$ values are given in the final column of 
Table~\ref{table:Compare}, 
along with their uncertainties 
(which are dominated by the measurement uncertainties
of column 2).
The listed values of $\Delta R_{\rm p}$ are almost all positive, and 
their weighted average is 
0.035(7)~fm 
(with an acceptable $\chi^2$ value of 7.3 for 14 degrees of freedom).
Thus, the atomic hydrogen data would be consistent with the 
theoretical predictions if $R_{\rm p}$ were 4\% larger than the 
CREMA value. 

This 4\% discrepancy has 
been the topic of great interest since the muonic measurement
of $R_{\rm p}$ was published \cite{pohl2010size}. 
A similar discrepancy has been found between 
the CREMA value for $R_{\rm p}$ and that obtained from 
electron proton scattering \cite{PRL.105.242001,bernauer2014electric,lee2015extraction}
(although our recent analysis \cite{horbatsch2015evaluation} of the scattering data does not
support the notion of this discrepancy).
The 4\% discrepancy for $R_{\rm p}$ 
is often referred to as the proton size puzzle,
and many works have discussed the puzzle 
(see Refs.~\cite{pohl2013muonic,bernauer2014proton,carlson2015proton} 
for reviews of this body of work). 

\section{summary}

We present clear formulas and tabulations of the 
bound-state energy levels for atomic hydrogen. 
The tabulation includes the new, more precise
value for the proton charge radius. 
The hyperfine structure corrections due to 
the anomalous moment and the finite mass of the 
proton, and due to off-diagonal matrix elements 
of the hyperfine Hamiltonian are clearly laid out,
along with general formulas for their dependence 
on $n$, $\ell$, $j$, and $f$. The energy of all bound states of atomic 
hydrogen can now be determined to a precision
of better than 1 kHz.

This work is supported by NSERC and CRC.

\bibliography{HydrogenEnergies}

\end{document}